\documentclass[12pt]{article}
\usepackage{array}
\begin{document}
\begin{flushright}
{\tt hep-ph/0311127}\\
OSU-HEP-03-13 \\
November, 2003 \\
\end{flushright}
\vspace*{2cm}
\begin{center}
{\baselineskip 25pt \large{\bf Model Building with Gauge--Yukawa
Unification  } \\


}

\vspace{1cm}

{\large
Ilia Gogoladze\footnote
{On a leave of absence from: Andronikashvili Institute of Physics, GAS, 380077, Tbilisi, Georgia.\\
email: {\tt ilia@hep.phy.okstate.edu}}$^{* \dagger}$, 
Yukihiro Mimura\footnote {email: {\tt mimura2y@uregina.ca}}$^{\ddagger } $ and 
S. Nandi\footnote {email:
{\tt shaown@okstate.edu}}$^\dagger$ } \vspace{.5cm}


{ $^*${\it Department of Physics, University of Notre Dame \\Notre
Dame, IN 46556, USA}\\
$^\dagger${\it Physics Department, Oklahoma State University, \\
             Stillwater, OK 74078}\\
$^{\ddagger  }${\it Department of Physics, University of Regina, \\
Regina, Saskatchewan S4S 0A2, Canada}
}
\vspace{.5cm}

\vspace{1.5cm}
{\bf Abstract}
\end{center}

In supersymmetric theories with extra dimensions, 
the Higgs and matter fields can be part of the gauge multiplet,
so that the Yukawa interactions can arise from the gauge interactions.
This leads to the possibility of  
gauge--Yukawa coupling unification, $g_{i}=y_{f}$,
in the effective four dimensional theory after the
initial gauge symmetry and the supersymmetry are broken upon orbifold compactification.
We consider gauge--Yukawa unified models based on a variety of
four dimensional symmetries, including $SO(10)$, $SU(5)$, Pati--Salam symmetry,
trinification, and the Standard Model.
Only in the case of Pati--Salam and the Standard Model symmetry,
we do obtain gauge--Yukawa unification. Partial gauge--Yukawa unification is
also briefly discussed.

\thispagestyle{empty}

\bigskip
\newpage

\addtocounter{page}{-1}

\section{Introduction}
\baselineskip 20pt

The Standard Model is well established experimentally
as an accurate description of physics below the weak scale.
Supersymmetry is one of the most promising candidate among models beyond the Standard Model.
Experimental data support the unification of the three gauge coupling constants 
at a grand unified scale within the Minimal Supersymmetric Standard Model (MSSM),
and the standard gauge group can be unified into a simple gauge group such as $SU(5)$.
Thus particle physicists have a paradigm
for the gauge sector in the Standard Model.
However, we have less theoretical understanding of the Yukawa sector.
In fact, most of the parameters in the Standard Model are in the Yukawa sector:
masses of quarks and leptons and their mixings.
Nonetheless, most theorists expect that we will understand how those parameters are determined
once we have a fully unified picture.

Recent work on theories in higher dimensions provides several interesting
phenomenological pictures.
Models exist in which the extra dimensions are compactified on an orbifold, 
such that the
transformation properties of the fields
are responsible for breaking unwanted symmetries, including gauge symmetries 
\cite{Scherk:1978ta, Hosotani:1983xw}.
%
Recently, Ref.\cite{Kawamura:1999nj} utilized
this orbifold breaking for $SU(5)$ grand unified
theory (GUT), and solved the doublet--triplet splitting problem by
projecting out the colored Higgs triplets using the orbifold
transformation properties. Since then orbifold GUTs 
have been
widely applied in building models in higher dimensions \cite{Kobakhidze:2001yk}.

Orbifold GUTs provide an interesting possibility:
The bulk Lagrangian can have symmetries which are broken on the four dimensional (4D) wall. 
Thus parameters in the 4D models can be related
even if the whole Lagrangian does not have exact symmetries.
For example, suppose that the bulk symmetry is $SO(10)$
which is broken down to Pati--Salam \cite{Pati:1974yy} symmetry,
$SU(4) \times SU(2)_L \times SU(2)_R$, in 4D.
There would be no reason for the three gauge couplings of the Pati--Salam
model to be related in the context of the 4D Lagrangian alone.
However, the three gauge couplings will in fact be unified due to the larger bulk symmetry.
Such a ``coincidence" would be very hard to justify from a 4D point of view, 
but is automatic in the higher dimensional theory.

Another attractive motivation for extending the dimensions
is to understand the variety of particles in nature
by means of a geometric language.
For example, in the original idea of Kaluza--Klein,
the 4D gauge fields are included in the higher dimensional metric tensor.
Gauge--Higgs unification \cite{Hosotani:1983xw,Manton:1979kb}
provides another attractive idea in higher dimensional theories.
Recent progress of the higher dimensional unified theories makes many people
revisit the idea of the gauge--Higgs 
unification \cite{Krasnikov:dt,Csaki:2002ur,Burdman:2002se,Haba:2002vc,Gogoladze:2003bb}.
The extra dimensional components of gauge fields transform
as scalar fields in 4D and
thus can be Higgs fields which break the 4D gauge symmetry.
Masses of these scalar fields are prohibited by gauge invariance,
and in supersymmetric theories
the scalar fields remain massless in the low energy
due to the non-renormalization theorem.
Thus these fields are good candidates
for the Higgs field of the Standard Model. 

In higher dimensional supersymmetric theories, the gauge
multiplet 
contains both vector and chiral superfields  
in the language of $N=1$ supersymmetry.
By assigning different transformation properties
to the vector and chiral superfields,
we can make the vector superfield massless while the chiral superfields become heavy,
breaking the extended supersymmetry down to $N=1$.
If we simultaneously break the gauge symmetry through orbifold
compactification, the chiral superfields which
correspond to the broken generators can have zero modes, which
remain massless in the low energy. 
Then, we can identify these fields
with the low energy fermions and Higgs fields.

Ref.\cite{Burdman:2002se} emphasizes an interesting
possibility that the gauge and Yukawa coupling constants have the
same origin, if matters (quarks and leptons) are also bulk fields
in the context of gauge--Higgs unification.  
Because the Yukawa interactions arise from gauge interactions in the higher dimensional Lagrangian,
the gauge and Yukawa coupling constants can be unified in the higher dimensional theory. 
This is also an example
that the parameters in 4D are related due to the
large bulk symmetry which we mentioned before. 
This fact can be a
strong motivation to consider the orbifold GUTs.
In Ref.\cite{Burdman:2002se}, the authors considered the 5D theories of 
$SU(3)_w$ and $SU(6)$ as an example of this scenario.
Gauge--Higgs unification in 5D theories with larger gauge groups
such as $E_6$, $E_7$ and $E_8$ has also been studied \cite{Haba:2002vc}. 
In Ref.\cite{Gogoladze:2003bb}, the authors
considered gauge--Higgs unification in $SU(4)_w$ and
$SO(12)$, and suggested that the orbifold breaking of
the $SU(4)_w$ gauge symmetry gives an economical realization of the
representations of the quarks and leptons. 
In Ref.\cite{Cosme:2003cq}, gauge--Higgs unification in $Sp(4)_w$
and $SO(11)$ were also suggested.

Another interesting possibility 
is that quarks and leptons can be unified in the gauge multiplet.
It is well known that three (or four) families of quarks and leptons can be contained
in the adjoint representations of large gauge groups, such as $E_7$ and $E_8$.
The matter in the adjoint representation is always vector-like,
but one can project out the vector-like partner
by a $Z_3$ transformation \cite{Babu:2002ti}.
This encourages us to consider the interesting possibility that the gauge and matter are
unified in higher dimensional models.
Alternatively one could consider that the three families originate
from chiral superfields in a gauge multiplet \cite{Watari:2002tf}
since the gauge multiplet in 6D $N=2$ supersymmetry
contains three $N=1$ chiral superfields in 4D.
The gauge and matter unification is also studied in Ref.\cite{Li:2003ee}.
The group theoretical aspects in such scenarios
were studied in Ref.\cite{Hebecker:2003jt}.

In Ref.\cite{Gogoladze:2003ci,Gogoladze:2003yw},
we suggested the possibility of
unifying gauge, matter and Higgs fields in one supersymmetric
gauge multiplet in higher dimensions, as well as the unification
of the gauge and Yukawa couplings. 
As a simple example, a 6D $N=2$
supersymmetric unified model was constructed.
The gauge symmetry group
$SU(8)$ or $SO(16)$ is broken down to the Pati--Salam group with
extra $U(1)$ symmetries in 4D through a $T^2/Z_6$ orbifold compactification,
and the theory is reduced to 4D $N=1$ supersymmetric Pati--Salam model.
The electroweak Higgs fields and Standard Model fermions
of the third family (two families in the $SO(16)$ model) are unified
with the gauge bosons in the 6D gauge multiplet. 
The 6D bulk gauge interaction produces Yukawa interactions,
which give masses
to the quarks and leptons by the Higgs mechanism,
giving rise to gauge--Yukawa unification.
This gauge--Yukawa unification is in good  numerical agreement with
experiment \cite{Gogoladze:2003ci},
and  predicts a large value for 
 $\tan \beta$, around $52$, and as well as interesting  relations among 
supersymmetric threshold corrections \cite{Gogoladze:2003pp}.

In this paper, we study gauge--Yukawa unified models
for various 4D symmetries.
We discuss  models  in 5D with $N=1$ supersymmetry compactified
on an $S^1/Z_2$ orbifold. 
We also discuss 6D models with a $Z_n$ gauge twist.
We consider the Pati--Salam gauge symmetry as well as the Standard
Model gauge symmetry in 4D, 
and the third family Yukawa couplings for quarks and leptons are unified to three 
gauge coupling constants of the Standard Model.
Models with 4D $SU(5)$ and $SO(10)$ symmetries 
are also constructed, but here the Yukawa couplings
are not necessarily unified to the gauge couplings even though the
Yukawa interactions do come from the gauge interactions. 
Models with the trinification gauge symmetry, $SU(3)_c \times SU(3)_L \times SU(3)_R$,
in 4D are also studied.

Our paper is organized as follows:
In section 2, we will present the idea of gauge--Yukawa unification.
We will see that Yukawa interactions can arise from gauge interactions 
if the Higgs fields are unified in the higher dimensional gauge multiplet 
and the quarks and leptons are zero modes of the bulk fields.
In section 3, we construct the several models with
gauge--Yukawa unification.
In section 4, we make remarks on the gauge--Yukawa unification.
Section 5 contains our discussions and conclusions.

\section{The Idea of Gauge--Yukawa Unification}

In this section, we will introduce 
the idea of gauge and Yukawa unification.
First of all, we discuss the gauge--Higgs unification in the higher
dimensional gauge theory. 
The extra dimensions are compactified on an orbifold space. 
Then, the gauge fields with the extra
dimensional coordinates behave as scalar fields in the 4D point of view.


Consider a 5D gauge theory on an $S^1/Z_2$ orbifold.
The 5D gauge fields are denoted as $A_M = (A_\mu,A_5)$,
where $A_\mu$ ($\mu=0,1,2,3$) are the usual 4D gauge bosons
and $A_5$ behaves as a scalar field in 4D.
The fifth dimensional coordinate $x_5$ is identified to $-x_5$
under the $Z_2$ parity,
and resulting extra dimensional space
is the interval $[0,\pi R]$.
The physical space has boundaries at $x_5 = 0$, $\pi R$.
The boundary condition for the 5D gauge fields is
\begin{equation}
A_\mu(-x_5) = A_\mu(x_5), \quad A_5(-x_5)= - A_5(x_5).
\end{equation}
Since the  $Z_2$ parity of $A_5$ is odd,
the zero mode of $A_5$ is projected out from mass spectrum.

We can break the gauge symmetry by orbifold boundary conditions \cite{Hebecker:2001jb}
\begin{equation}
A_\mu(-x_5) = P A_\mu(x_5) P^{-1}, \quad A_5(-x_5)= - P A_5(x_5)P^{-1},
\end{equation}
where $P$ acts on the gauge space as an inner automorphism with $P^2$ to be identity.
With this boundary condition, the gauge group $G$ is broken down to a subgroup $H$.
We denote the generator of the gauge group $G$ as $T^A$, that of subgroup $H$ as $T^a$;
thus $\{T^A\} = \{T^a\} \oplus \{T^{\hat a}\}$.
The $Z_2$ parity of $A^a_\mu$ is even, while that of $A^{\hat a}_\mu$ is odd.
The zero modes of odd fields are projected out, and
thus the gauge symmetry $G$ is broken to $H$.
At the same time, the $Z_2$ parities of $A_5$ are opposite to those of $A_\mu$,
and therefore, $A_5$ with broken generators have zero modes.
The vacuum expectation values of $A_5^{\hat a}$ can break the symmetry $H$.
For example, in the case where $G=SU(3)$ and $H= SU(2) \times U(1)$,
the broken generators $T^{\hat a}$ correspond to two $SU(2)$ doublets:
\begin{equation}
\mathbf{8} = \overbrace{\mathbf{3}_0 + \mathbf{1}_0}^{T^a} +
\overbrace{\mathbf{2}_1 + \mathbf{2}_{-1}}^{T^{\hat a}}.
\end{equation}
Thus, vacuum expectation values of $A_5^{\hat a}$ breaks $H$ down to $U(1)^\prime$.
Of course, since the Higgs fields must be complex and
quartic Higgs couplings are needed,
we have to consider the supersymmetric version of the 5D models or 6D theories
in order to identify the doublets as Standard Model Higgs fields.

Suppose that $A_5^{\hat a}$ can be identified with (the imaginary part of) 
the  Higgs fields,
and quarks and/or leptons are zero modes of higher dimensional fermions.
The 5D fermion $\psi_5$ contains both left- and right-chiral
4D fermions, $\psi_5 = (\psi_L, \psi_R^c)$.
4D zero modes can be made chiral through the orbifold boundary condition.
The kinetic term of the fermion is written as
$i \bar \psi_5 \Gamma^M D_M \psi_5$, where $\Gamma_M$ is 5D gamma matrices
$\Gamma_M = (\gamma_\mu, i\gamma_5)$ and $D_M$ is the gauge covariant derivative.
The kinetic term is written in 4D form:
\begin{equation}
i \bar \psi_5 \Gamma^M D_M \psi_5 = i \overline{\psi_L} \gamma^\mu D_\mu \psi_L +
i \overline{\psi_R} \gamma^\mu D_\mu \psi_R
+ (\overline{\psi_R} D_5 \psi_L + {\rm h.c.}).
\end{equation}
If the fermion $\psi_L$ is in the fundamental representation of the gauge group $G$,
then the last term can be written as
\begin{equation}
\label{mm5}
\overline{\psi_R} \partial_5 \psi_L - i g \overline{\psi_R} A_5 \psi_L + {\rm h.c.}
\end{equation}
The first term in Eq.(\ref{mm5}) gives mass terms of Kaluza--Klein excited states,
and the second term can be interpreted as a Yukawa
interaction which gives masses to 4D zero modes of quarks and leptons. 

We now consider its supersymmetric version in order to build more realistic models.
5D $N=1$ supersymmetry corresponds to 4D $N=2$ supersymmetry.
The $N=2$ gauge multiplet contains both an $N=1$ vector multiplet $V(A_\mu,\lambda)$
and a chiral multiplet $\Sigma$ whose scalar component is $(\sigma + i A_5)/\sqrt2$.
The orbifold boundary conditions are given as
\begin{equation}
V(-x_5) = P V(x_5) P^{-1}, \quad \Sigma(-x_5) = -P\Sigma(x_5)P^{-1},
\label{bc:n=2gauge}
\end{equation}
and the $N=2$ supersymmetry is broken to $N=1$ supersymmetry upon compactification to 4D.
The quarks and leptons can be included in $N=2$ hypermultiplets
which contains $N=1$ chiral multiplets $(\Psi,\Psi^c)$.
The supersymmetric version of the term that leads to Yukawa interactions
is
\begin{equation}
\int d^2 \theta (\Psi^c \partial_5 \Psi- \sqrt2 g \Psi^c \Sigma \Psi) + {\rm h.c.}
\end{equation}
assuming the field $\Psi$ is in the fundamental representation of the gauge group $G$.
The gauge transformations for the chiral multiplets are given as
\begin{equation}
\Sigma \rightarrow e^{\Lambda} (\Sigma - \frac1{\sqrt2 g} \partial_5) e^{-\Lambda},
\quad \Psi \rightarrow e^{\Lambda} \Psi, \quad \Psi^c \rightarrow \Psi^c e^{-\Lambda},
\end{equation}
under which the interaction term is invariant. 
The orbifold boundary condition for hypermultiplet is
\begin{equation}
\Psi (-x_5) = P \Psi(x_5), \quad \Psi^c (-x_5) = - \Psi^c (x_5)  P^{-1},
\label{bc:n=2hyper}
\end{equation}
and this makes the 4D theory chiral.
The quartic coupling of $\Sigma$ arises from the $D$-term of the supersymmetric Lagrangian.
If the zero modes of chiral field $\Sigma$ can be identified with the Standard Model Higgs fields
and the zero modes of $\Psi$ and $\Psi^c$ contain left- and right-chiral matters, then 
the gauge interaction term $\sqrt2 g \Psi^c \Sigma \Psi$ includes the
Yukawa interactions in 4D \cite{Burdman:2002se}.
Several models have been  realized including models based on
$SU(3)_w$  \cite{Burdman:2002se},
$SU(4)_w$  \cite{Gogoladze:2003bb},
$Sp(4)_w$  \cite{Cosme:2003cq},
and for grand unified models, $SU(6)$ \cite{Burdman:2002se},
$SO(11)$ \cite{Cosme:2003cq}, $SO(12)$ \cite{Gogoladze:2003bb},
$E_6$, $E_7$, and $E_8$ \cite{Haba:2002vc}.

We can easily extend the gauge--Yukawa unification to 
6D $N=2$ supersymmetric models on a $T^2/Z_n$ orbifold \cite{Gogoladze:2003ci}.
6D $N=2$ supersymmetry corresponds to 4D $N=4$ supersymmetry. 
The $N=4$ gauge multiplet contains an $N=1$ vector multiplet $V$ and three
$N=1$ chiral multiplets $\Sigma$, $\Phi$ and $\Phi^c$ in the
adjoint representation. 
The scalar component of $\Sigma$ is $(A_6 + i A_5)/\sqrt2$.
It is important to notice that 6D bulk field is only the gauge multiplet.
The $T^2/Z_n$ orbifold is constructed by identifying
extra dimensional complex coordinate $z\rightarrow \omega z$, where
$\omega^n=1$. The number $n$ is restricted to $n=2,3,4,6$ in the
case of toroidal compactification.
The orbifold conditions are given as
\begin{eqnarray}
V(\omega z, \bar \omega \bar z) &=& {\cal R}\cdot V(z,\bar z), \label{twist:1}\\
\Sigma(\omega z, \bar \omega \bar z) &=& \bar \omega \ {\cal R} \cdot \Sigma(z,\bar z), \label{twist:2} \\
\Phi(\omega z, \bar \omega \bar z) &=& \omega^l \ {\cal R} \cdot \Phi(z,\bar z), \\
\Phi^c(\omega z, \bar \omega \bar z) &=& \omega^m \ {\cal R} \cdot \Phi^c(z,\bar z), \label{twist:4}
\end{eqnarray}
where ${\cal R}$ acts on the gauge space and ${\cal R}^n$ is the identity mapping.
The number $l$ and $m$ have a relation $l+m=1$ (mod $n$).
In the case where $n>2$, the $N=4$ supersymmetry is broken down to $N=1$ supersymmetry
in 4D through the above conditions.
The gauge interaction term which includes Yukawa interaction is given as
\begin{equation}
\int d^2 \theta \:{\rm Tr} \,\frac1k \,(\Phi^c \partial \Phi - \sqrt2 g \,\Phi^c [\Sigma, \Phi]) +{\rm h.c.}
\end{equation}
where $\partial = \partial_5 - i\partial_6$ and $k$ is a normalization factor for group generators.
This is invariant under gauge transformation
\begin{equation}
\Sigma \rightarrow e^{\Lambda} (\Sigma - \frac1{\sqrt2 g} \partial_5) e^{-\Lambda},
\quad \Phi \rightarrow e^{\Lambda} \Phi e^{-\Lambda},
\quad \Phi^c \rightarrow e^{\Lambda} \Phi^c e^{-\Lambda}.
\end{equation}
If the left- and right-chiral matter and MSSM Higgs fields are contained
in the zero modes of the chiral multiplets $\Sigma$, $\Phi$ and $\Phi^c$,
the gauge interaction term, Tr $\Phi^c [\Sigma, \Phi]$, includes the Yukawa interactions.
Since Tr $\Phi^c [\Sigma, \Phi]$ term
can be written by using group structure functions, the Yukawa interaction can be determined
by the group structure functions.
In this construction, the matter, Higgs and gauge fields can be unified in a single multiplet.
This idea has been realized in models with $SU(8)$ or $SO(16)$ bulk symmetries
\cite{Gogoladze:2003ci,Gogoladze:2003yw}.
The $Z_n$ twisted boundary conditions (\ref{twist:1}--\ref{twist:4})
can be also considered in the two dimensional disk space $D^2/Z_n$ and annulus space
$A^2/Z_n$ \cite{Li:2001dt} and in a conifold construction \cite{Hebecker:2003jt}.

The idea of gauge--Higgs unification can also be realized in a
non-supersymmetric model in 6D \cite{Csaki:2002ur}.
But, there are some reasons to prefer the supersymmetric version.
It is well known that supersymmetry can solve the gauge hierarchy problem.
Though supersymmetry can
solve the hierarchy problem, there is no reason that the
supersymmetric Higgs mass should be very light compared to GUT scale. It
is considered that there is a symmetry to prohibit the large
supersymmetric Higgs mass. In the context of gauge--Higgs
unification, it is gauge invariance that prohibits the large supersymmetric Higgs mass.
Another advantage for
supersymmetric gauge--Higgs unification is the possibility that the
third family Yukawa couplings 
might unify with gauge couplings 
at a grand unified scale \cite{Gogoladze:2003ci, Gogoladze:2003pp}.
Thus gauge--Higgs unification may present a new paradigm for understanding
the origin of families and the structure of quark and lepton mass matrices. Of
course, the idea of gauge--Yukawa unification can be applied to
more than 6D models since the interactions are simply
in the covariant derivative terms in higher dimensional
Lagrangian.

\section{Unification of the Gauge and Yukawa Couplings }

If the Higgs fields do originate from the gauge multiplet,
and the left- and right-chiral matters are bulk fields,
then the Yukawa interaction can arise from gauge interactions
as we have discussed in previous section.
In this section,
we will discuss about the unification of gauge and third-family Yukawa couplings.


If we take into account the kinetic normalization
of the 4D zero modes,
we find that the Yukawa couplings arising from the bulk gauge interaction
is just same as conventional gauge couplings.
Thus, if the effects of brane-localized
interaction and the threshold corrections are  small enough,
the gauge and Yukawa couplings can be unified at
grand unified scale unless the family is largely mixed with other fields.
Therefore, we will first consider only the bulk interaction as zero-th order
approximation,
and see whether the prediction is viable for the various 4D symmetry.

\subsection{$SO(10)$}

The gauge--Higgs unification for
$SO(10)$ model in 4D is realized by considering the
5D bulk symmetry $SO(12)$ (or $SO(11)$).
%
%
The adjoint representation of $SO(12)$, $\mathbf {66}$-plet,
is decomposed under $SO(10) \times U(1)$ as
\begin{equation}
\mathbf{66} = \mathbf{45}_0 + \mathbf{1}_0 + \mathbf{10}_2 + \mathbf{10}_{-2},
\end{equation}
where two $\mathbf{10}$'s correspond to broken generators.
As we have seen in previous section, a 5D gauge multiplet contains
both a vector supermultiplet $V$ and a chiral supermultiplet $\Sigma$.
The boundary condition Eq.(\ref{bc:n=2gauge}) breaks $SO(12)$ symmetry,
and decomposed gauge multiplets $V_{\mathbf{45}_0}$ and $V_{\mathbf{1}_0}$ 
have massless modes, while $V_{\mathbf{10}_2}$ and $V_{\mathbf{10}_{-2}}$
do not have massless modes.
On the other hand, chiral multiplets $\Sigma_{\mathbf{10}_2}$ and 
$\Sigma_{\mathbf{10}_{-2}}$ have massless modes
and the zero modes can be identified with Higgs fields
which contain Standard Model Higgs doublets.

The matter fields are contained in the hypermultiplet $(\Psi,\Psi^c)$, 
and the chiral supermultiplet $\Psi$ is in the spinor representation, $\mathbf{32}$.
Although the representation of $SO(12)$ is real ($\mathbf{\overline{32} = 32}$), 
the orbifold boundary condition for hypermultiplet Eq.(\ref{bc:n=2hyper})
makes the 4D theory chiral.
The spinor representation is decomposed under $SO(10) \times U(1)$ as
\begin{equation}
\mathbf{32} = \mathbf{16}_1 + \mathbf{\overline{16}}_{-1}.
\end{equation}
The $Z_2$ parities of the hypermultiplet $(\Psi, \Psi^c)$ are assigned as
\begin{equation}
\Psi_\mathbf{32} = \Psi_{\mathbf{16}_1}^{(+)} + \Psi_{\overline{\mathbf{16}}_{-1}}^{(-)},
\qquad
\Psi^c_\mathbf{32} = \Psi_{\overline{\mathbf{16}}_{-1}}^{c\ (-)} + \Psi_{\mathbf{16}_{1}}^{c\ (+)},
\end{equation}
where the $Z_2$ parities are denoted in the superscripts,
and one hypermultiplet has two $\mathbf{16}$-dimensional
chiral multiplets as the 4D zero modes. We can
identify the zero modes as two families of matters. 
Then the bulk gauge
interaction involving the hypermultiplet and the gauge multiplet
includes the Yukawa interaction
\begin{equation}
\Psi^c \Sigma \Psi \supset \Psi_{\mathbf{16}_1}  \Sigma_{\mathbf{10}_{-2}}  \Psi^c_{\mathbf{16}_{1}}
\label{SO(10)_int}
\end{equation}
However, this bulk Yukawa interaction is not viable for phenomenology
since two eigenmasses for the two families are degenerate.
In order to break degeneracy, we have to introduce
$\Psi_\mathbf{16} \Psi_\mathbf{16} \Sigma_\mathbf{10}$ or
$\Psi^c_\mathbf{16} \Psi^c_\mathbf{16} \Sigma_\mathbf{10}$ terms as the brane
interactions. 
However, since we are assuming that the brane-localized interaction should
be less than bulk interaction, such brane-localized terms cannot
realize the fermion mass hierarchy. 
Thus we have to introduce an extra brane field $X_{\overline{\mathbf{16}}}$,
and 
the brane interaction such as
\begin{equation}
(M_1 \Psi_\mathbf{16} + M_2 \Psi^c_\mathbf{16}) X_\mathbf{\overline{16}}.
\end{equation}
Then
one of the combination of two $\mathbf{16}$'s becomes heavy, and the
Yukawa coupling for another combination is proportional to the
mixing ${2 M_1 M_2}/{(M_1^2 + M_2^2)}$. Thus, in general, the
Yukawa couplings for the third family are not equal to gauge
couplings in this model.
So, in this $SO(12)$ model, the gauge--Higgs unification is realized
and the Yukawa interaction can arise from gauge interaction,
but
the unification of gauge--Yukawa couplings is not realized in general.

We can construct gauge--Higgs--matter unification in $SO(10)$ model
by considering 6D bulk $E_7$ symmetry
and using the following breaking chain:
\begin{equation}
E_7 \rightarrow SO(12) \times SU(2) \rightarrow SO(10) \times SU(2) \times U(1).
\end{equation}
The adjoint representation of $E_7$ is decomposed under $SO(10) \times SU(2) \times U(1)$ as
\begin{eqnarray}
\mathbf{133} &=& (\mathbf{66,1}) + (\mathbf{1,3}) + (\mathbf{32',2}) 
\qquad (\mbox{under } SO(12) \times SU(2)) \\
&=& (\mathbf{45,1})_0 + (\mathbf{1,1})_0 + (\mathbf{10,1})_2 + (\mathbf{10,1})_{-2}
    + (\mathbf{1,3})_0
    + (\mathbf{16,2})_{-1} + (\mathbf{\overline{16},2})_1 \ .
\end{eqnarray}
We assign the $Z_n$ charge to the vector multiplet $V$ as
\begin{equation}
\begin{array}{|c||c|c|c|c|c|c|c|}
\hline
& V_{(\mathbf{45,1})_0} & V_{(\mathbf{1,1})_0} & V_{(\mathbf{1,3})_0} &
V_{(\mathbf{10,1})_2} & V_{(\mathbf{10,1})_{-2}}
& \ V_{(\mathbf{16,2})_{-1}} & V_{(\mathbf{\overline{16},2})_1}  \\ \hline
Z_n & 0 & 0 & 0 & x & -x & y & -y  \\
\hline
\end{array}
\end{equation}
where $x$ and $y$ is not zero (mod $n$).
The automorphism condition (${\rm Tr} \Sigma_{\mathbf{(10,1)}} [\Phi_{\mathbf{(16,2)}},\Phi^c_{\mathbf{(16,2)}}]$
is invariant) is $2y + x \equiv 0$ (mod $n$), where $\Sigma$, $\Phi$, and $\Phi^c$ are chiral multiplets
in the gauge multiplet.
We can break $SU(2)$ gauge symmetry by assigning $Z_n$ charge of $V_{(\mathbf{16,2})}$
as $(y_1,y_2)$ instead of $(y,y)$.
In that case, the automorphism condition is $y_1 + y_2 + x \equiv 0$ (mod $n$).
Then, extracting zero modes from chiral fields $\Sigma$, $\Phi$, $\Phi^c$ appropriately,
we find that the bulk interaction includes Yukawa couplings for zero modes,
$\mathbf{16}_1 \mathbf{16}_2 \mathbf{10}$,
where the subscripts denote $SU(2)$ indices.
For instance, by choosing $y_1 \equiv 1$, $y_2 \equiv -l$, $x \equiv -m$,
we find that $\mathbf{16}_1$, $\mathbf{16}_2$, and $\mathbf{10}$ lie in the
chiral multiplets $\Sigma$, $\Phi$ and $\Phi^c$, respectively,
from the $Z_n$ transformation of their chiral multiplets (\ref{twist:2}--\ref{twist:4}).
In such a way, both Higgs and fermion fields are unified in the gauge multiplet.  
However, the fermion masses are also degenerate since bulk
gauge interaction does not include flavor diagonal Yukawa interactions.
Thus the gauge and Yukawa coupling constants are not unified in general
in the same way as 5D $SO(12)$ model.

We can also construct gauge--Higgs--matter unification in $SO(10)$ model
by considering 6D bulk $E_8$ symmetry
and using the following breaking chain:
\begin{equation}
E_8 \rightarrow SO(16) \rightarrow SO(10) \times SU(4).
\end{equation}
The adjoint representation of $E_8$ is decomposed under $SO(10) \times SU(4)$ as
\begin{eqnarray}
\mathbf{248} &=& \mathbf{120} + \mathbf{128} 
\qquad (\mbox{under } SO(16)) \\
&=& (\mathbf{45,1}) + (\mathbf{1,15}) + (\mathbf{10,6}) + (\mathbf{16,\bar 4})
    + (\mathbf{\overline{16},4}).
\end{eqnarray}
As in the case of $E_7$ model, gauge--Higgs--matter unification can be realized,
but the gauge and Yukawa couplings  are not unified in general.

\subsection{Pati--Salam: $SU(4)_c \times SU(2)_L \times SU(2)_R$}

The reason why the eigenmasses of two families 
are degenerate in $SO(10)$ in the previous subsection 
is that both left- and right-chiral
matters are unified in one multiplet. The problem can be solved by
splitting the left--right matters by orbifold conditions. 
So, for example, it is good to consider Pati--Salam symmetry as a 
4D gauge symmetry to obtain gauge--Yukawa unified models. 

The $SO(10)$ representations $\mathbf{16}$ and $\mathbf{10}$
are decomposed under Pati--Salam symmetry as
\begin{equation}
\mathbf{16} = L_\mathbf{(4,2,1)} + \bar R_\mathbf{(\bar 4,1,2)}^\prime,
\quad
\mathbf{16}^\prime = L^\prime_\mathbf{(4,2,1)} + \bar R_\mathbf{(\bar 4,1,2)},
\quad
\mathbf{10} = H_\mathbf{(1,2,2)} + C_\mathbf{(6,1,1)},
\end{equation}
and the $SO(10)$ interaction $\mathbf{16}\cdot\mathbf{10}\cdot\mathbf{16^\prime}$
is written in terms of the Pati--Salam representations as
\begin{equation}
\mathbf{16} \cdot \mathbf{10} \cdot \mathbf{16}^\prime
= L H \bar R + \bar R^\prime H L^\prime + L C L^\prime + \bar R^\prime C \bar R.
\end{equation}
The interaction includes Yukawa interaction for two families with degenerate
Yukawa couplings.
However,
if $\bar R^\prime$, $L^\prime$ (and $C$) are projected out by orbifold,
the bulk gauge interaction gives the zero-mode Yukawa interaction
for only one family which we can identify as the third family.
The projection can be realized by considering the $SO(12)$ breaking chain:
\begin{equation}
SO(12) \rightarrow \left\{
           \begin{array}{c}
              SO(10) \times U(1) \\
              SU(4) \times SU(4) 
           \end{array}
                   \right\}
\rightarrow SU(4) \times SU(2) \times SU(2) \times U(1).
\end{equation}
The gauge--Higgs unification of
Pati--Salam models in 5D $SO(12)$ bulk symmetry is
considered in Ref.\cite{Gogoladze:2003bb}.

The gauge--Higgs--matter unified models in Pati--Salam model starting
from 6D $SU(8)$ or $SO(16)$ bulk symmetries are
studied in the Ref.\cite{Gogoladze:2003ci,Gogoladze:2003yw}.
The $SU(8)$ symmetry is broken by orbifold condition as
\begin{equation}
SU(8) \rightarrow SU(4) \times SU(2) \times SU(2) \times U(1)^2,
\end{equation}
and the adjoint representation of $SU(8)$ is decomposed as
\begin{equation}
\mathbf{63} = \left(
\begin{array}{ccc}
\mathbf{(15,1,1)}_{0,0} & \mathbf{(4,2,1)}_{2,0} & \mathbf{(4,1,2)}_{2,4} \\
\mathbf{(\bar 4,2,1)}_{-2,0} & \mathbf{(1,3,1)}_{0,0} & \mathbf{(1,2,2)}_{0,4} \\
\mathbf{(\bar 4,1,2)}_{-2,-4} & \mathbf{(1,2,2)}_{0,-4} & \mathbf{(1,1,3)}_{0,0}
\end{array}
\right) + \mathbf{(1,1,1)}_{0,0} + \mathbf{(1,1,1)}_{0,0},
\end{equation}
where the subscripts denote charges under the $U(1)_1 \times U(1)_2$ symmetry.
The trilinear bulk interaction includes the
one-family Yukawa coupling
\begin{equation}
(\mathbf{4,2,1})_{2,0}\  (\mathbf{\bar 4,1,2})_{-2,-4}\ (\mathbf{1,2,2})_{0,4}.
\end{equation}

In the 6D bulk $SU(8)$ or $SO(16)$ models,
we obtain the unification condition for the gauge and the Yukawa couplings
at grand unified scale:
\begin{equation}
g_3 = g_2 = g_1 = y_t = y_b = y_\tau = y_{\nu_\tau}^{\rm (Dirac)},
\end{equation}
assuming that Pati--Salam symmetry is broken down to the Standard Model 
by the Higgs mechanism at grand unified scale.
Such an unification of the gauge and Yukawa coupling constants is quite an interesting
possibility.
In the case where 4D symmetry is $SO(10)$,
the relation between gauge and Yukawa coupling constants
still has ambiguity as we have seen in the last subsection,
though the Yukawa interaction arises from the gauge interaction.
It is interesting that the unification of the gauge and the Yukawa couplings
can be achieved in 4D Pati--Salam symmetry.

\subsection{$SU(5)$}

The gauge--Higgs unification of $SU(5)$ model in 4D is
realized by considering 5D bulk symmetry $SU(6)$,
broken down to $SU(5) \times U(1)$ upon compactification on an
$S^1/Z_2$ orbifold. 
The gauge multiplet, adjoint representation of $SU(6)$,
contains both a vector multiplet $V$ and a chiral multiplet $\Sigma$. 
The adjoint representation is decomposed under $SU(5) \times U(1)$ as
\begin{equation}
\mathbf{35} = \mathbf{24}_0 + \mathbf{1}_0 + \mathbf{5}_6 + \overline{\mathbf{5}}_{-6}
\end{equation}
and the broken generator corresponds to $\mathbf{5}$ and $\overline{\mathbf{5}}$,
and $\Sigma_{\mathbf{5}}$ and $\Sigma_{\mathbf{\bar 5}}$
can be identified with the Higgs fields in $SU(5)$ model.

The matter fields are contained in the hypermultiplets in the $SU(6)$ representations
$\mathbf{15}$ and $\mathbf{20}$.
To make $\mathbf{10}$ and $\bar \mathbf{5}$ massless modes,
we assign the $Z_2$ parity of hypermultiplets as
\begin{equation}
\Psi_\mathbf{15} = \Psi_{\mathbf{10}_2}^{(+)} + \Psi_{\mathbf{5}_{-4}}^{(-)},
\qquad
\Psi^c_\mathbf{15} = \Psi_{\overline{\mathbf{10}}_{-2}}^{c\ (-)} + 
\Psi_{\overline{\mathbf{5}}_{4}}^{c\ (+)},
\end{equation}
\begin{equation}
\Psi_\mathbf{20} = \Psi_{\mathbf{10}_{-3}}^{(+)} + \Psi_{\overline{\mathbf{10}}_{3}}^{(-)},
\qquad
\Psi^c_\mathbf{20} = \Psi_{\overline{\mathbf{10}}_{3}}^{c\ (-)} + 
\Psi_{{\mathbf{10}}_{-3}}^{c\ (+)},
\end{equation}
and then we have three $\mathbf{10}$'s and one $\overline{\mathbf{5}}$
as 4D zero modes. 
The bulk gauge interaction
includes the Yukawa interaction
\begin{equation}
\label{mm} 
\Psi^c \Sigma \Psi \supset \Psi_{\mathbf{10}_{-3}} 
\Sigma_{\mathbf{5}_{6}}  \Psi_{\mathbf{10}_{-3}}^c 
+ \Psi_{\mathbf{10}_2} \Sigma_{\overline{\mathbf{5}}_{-6}} \Psi^c_{\overline{\mathbf{5}}_4}.
\end{equation}

Again, we find that the first term includes two-family up-type
quark Yukawa couplings
and the two eigenmasses for up-type quarks are degenerate.
So we have to introduce brane-localized interaction in the same way as in the $SO(10)$ case.
In this case, actually, we need to introduce two $\overline{\mathbf{10}}$'s
(or $\overline{\mathbf{5}}$'s) to cancel brane-localized gauge anomaly.
We call the brane fields as $X_{\overline{\mathbf{10}}}$ and $Y_{\overline{\mathbf{10}}}$.
The brane-localized interaction is written as
\begin{equation}
(h_1 S_1 \Psi_{\mathbf{10}_{-3}} + h_2 S_2 \Psi_{\mathbf{10}_{-3}}^c + h_3 S_3 \Psi_{\mathbf{10}_2})
X_{\overline{\mathbf{10}}}
+
(h_1^\prime S_1 \Psi_{\mathbf{10}_{-3}} + h_2^\prime S_2 \Psi_{\mathbf{10}_{-3}}^c 
+ h_3^\prime S_3 \Psi_{\mathbf{10}_2})
Y_{\overline{\mathbf{10}}},
\end{equation}
where $S_i$'s are $SU(5)$ singlet fields with appropriate $U(1)$ charges
and $h$'s are coupling constants,
and then one linear combination of three $\mathbf{10}$'s are massless.
Due to the mixing of three $\mathbf{10}$'s,
not only the up-type Yukawa but also the down-type Yukawa couplings
are not unified with the gauge couplings in general.

Next we construct gauge--Higgs--matter unification in $SU(5)$ model by considering 6D
bulk $E_7$ symmetry.
One of the regular maximal subgroup of $E_7$ is $SU(8)$ and the $SU(8)$ has
subgroup $SU(5)\times SU(3) \times U(1)$.
The adjoint representation of $E_7$ is decomposed under $SU(5) \times SU(3) \times U(1)$ as
\begin{eqnarray}
\mathbf{133} &=& \mathbf{63} + \mathbf{70} \qquad ({\rm under} \ SU(8)) \\
&=&  (\mathbf{24,1})_0 + (\mathbf{1,8})_0 + (\mathbf{1,1})_0 +
(\mathbf{5,\bar 3})_2 + (\mathbf{\bar 5,3})_{-2} \nonumber \\
&& + \ (\mathbf{\bar 5,1})_3 + (\mathbf{\overline{10},3})_1 + (\mathbf{10,\bar 3})_{-1}
+ (\mathbf{5,1})_{-3}\ .  
\label{branch}
\end{eqnarray}
%
We assign the $Z_n$ charge to vector multiplet $V$ as
\begin{equation}
\begin{array}{|c||c|c|c|c|c|c|c|c|c|}
\hline
& V_{(\mathbf{24,1})_0} & V_{(\mathbf{1,8})_0} & V_{(\mathbf{1,1})_0} &
V_{(\mathbf{5,\bar 3})_2} & V_{(\mathbf{\bar 5,3})_{-2}}
& V_{(\mathbf{\bar 5,1})_3} & V_{(\mathbf{\overline{10},3})_1} & V_{(\mathbf{10,\bar 3})_{-1}}
& V_{(\mathbf{5,1})_{-3}} \\ \hline
Z_n & 0 & 0 & 0 & -a & a & -z & -z+a & z-a & z \\
\hline
\end{array}
\end{equation}
and $a,z,z-a$ is not zero modulo $n$. Then the $E_7$ symmetry is broken down to
$SU(5) \times SU(3) \times U(1)$.
The automorphism condition ($[(\mathbf{10,\bar 3}), (\mathbf{10,\bar 3})]\
(\mathbf{5,\bar 3}) \subset \mathbf{[70,70]\ 63}$ is $Z_n$ invariant)
is given as $2z - 3 a \equiv 0$ (mod $n$).

This branch of $E_7$ is interesting since all the matters and Higgs
quantum numbers are included in the adjoint representation of 
$E_7$ : matter fields with three families,
$(\mathbf{10,\bar 3})$, $(\mathbf{\bar 5,3})$
and Higgs fields $(\mathbf{5,1})$, $(\mathbf{\bar 5,1})$.
However, in the sense of gauge--Yukawa unification,
such interpretation is not good because up-type Yukawa coupling
is not included in the bulk interaction.
In fact, $SU(3)$ symmetry prohibits the renormalizable up-type Yukawa coupling
in such scenario.
Thus, we adopt another interpretation of Higgs representations
such that the up-type Higgs field is in the $(\mathbf{5,\bar 3})$.
Then the bulk interactions include
\begin{equation}
(\mathbf{10,\bar 3})\ (\mathbf{10,\bar 3})\ (\mathbf{5,\bar 3}) +
(\mathbf{10,\bar 3})\ (\mathbf{\bar 5,3})\ (\mathbf{\bar 5,1}),
\end{equation}
and give rise to both up- and down-type Yukawa couplings.
%

We can also consider the branch $SU(6) \times SU(3)$ as a regular
maximal subgroup of $E_7$.
The decomposition of $E_7$ adjoint under the branch
is
\begin{equation}
\mathbf{133} = (\mathbf{35,1}) + (\mathbf{1,8}) + (\mathbf{15,\bar 3}) + (\mathbf{\overline{15},3}),
\end{equation}
and the $SU(6)$ is broken down to $SU(5) \times U(1)$.
Then we can obtain same decomposition as Eq.(\ref{branch}).
The up- and down-type Yukawa couplings are included in $(\mathbf{15,\bar 3})^3$ and
$(\mathbf{15,\bar 3})(\mathbf{\overline{15},3})(\mathbf{35,1})$, respectively.

We will break $SU(3)$ symmetry by assigning $Z_n$ charge of $(\mathbf{\bar 5,3})$
as $(a_1,a_2,a_3)$ instead of $(a,a,a)$.
The automorphism condition becomes $2z - a_1 - a_2 - a_3 \equiv 0$ (mod $n$).
Then extracting the zero modes
appropriately,
 we find that the bulk interaction includes the zero-modes interactions
\begin{equation}
\mathbf{10}_1 \mathbf{10}_2 \mathbf{5}_3 + \mathbf{10}_1 \mathbf{\bar 5}^1 \mathbf{\bar 5},
\end{equation}
where the scripts denote $SU(3)$ indices.
For instance,
suppose that $z-a_1 \equiv 1$, $z-a_2 \equiv -l$, $-a_3 \equiv -m$,
we find that $\mathbf{10}_1$, $\mathbf{10}_2$ and $\mathbf{5}_3$
lie in the chiral multiplets $\Sigma$, $\Phi$, $\Phi^c$, respectively, from their
$Z_n$ transformations (\ref{twist:2}--\ref{twist:4}).
Then up-type Yukawa coupling arises from bulk interaction.
To have down-type Yukawa coupling in the bulk interaction,
we need to have conditions that $a_1\equiv -m$, $-z \equiv -l$,
then $\mathbf{\bar 5}^1$, $\mathbf{\bar 5}$ lie in the $\Phi^c$ and $\Phi$
respectively.
We have another condition $a_2 \equiv 2z$ to satisfy above.
We note that
if we assume that $\mathbf{\bar 5}^1$, $\mathbf{\bar 5}$ lie in the $\Phi$ and $\Phi^c$,
we can see that the unbroken symmetry is not $SU(5)$ but $SO(10)$
since the condition at that time is $z-a_3\equiv 0$
which means the vector multiplets for $\mathbf{10}_3$ and $\mathbf{\overline{10}}^3$
are massless.

In this case, we also encounter the problem that two eigenmasses are degenerate for up-type quarks,
and we have to introduce the interaction such as
$(S_1 \mathbf{10}_1 + S_2 \mathbf{10}_2) \mathbf{\overline{10}}$,
where $S_1$ and $S_2$ are singlet under $SU(5)$ with suitable $U(1)$ charges.
This interaction can be bulk interaction in this model.

We note that we can make assignment for flipped-type $SU(5)$
since $E_7$ has a subgroups of $SU(5)\times U(1)$ 
with proper $U(1)$ charges for the flipped $SU(5)$.

\subsection{Standard Model Gauge Group : $SU(3)_c \times SU(2)_L \times U(1)_Y$}

The gauge--Higgs unification of the Standard Model in 5D
$SU(6)$ bulk symmetry is considered in Ref.\cite{Burdman:2002se}.
In their model, Yukawa coupling arises
from gauge interaction, but the gauge--Yukawa coupling unification
is not realized. In this subsection, we will consider the model
with 6D $E_7$ bulk symmetry which we have suggested in the
previous subsection. The $SU(5)$ symmetry is also broken by $Z_n$
twist in the model. Then we realize the unification of the
gauge and Yukawa couplings in the Standard Model. The reason is
basically same as the explanation which we gave in the case of
Pati--Salam symmetry: The left- and right-handed matters in
$\mathbf{10}$ are split by orbifold conditions.

As we have seen, $E_7$ has maximal subgroup $SU(5) \times SU(3)_H \times U(1)_X$.
We break the $SU(5)$ symmetry down to $SU(3)_c \times SU(2)_L \times U(1)_Y$
by orbifold compactification. 
We assign $Z_n$ charges to $SU(5)$ vector multiplet, $V_\mathbf{24}$, as
\begin{equation}
\begin{array}{|c||c|c|c|c|c|} \hline
 & V_{(\mathbf{8,1})_0} & V_{(\mathbf{1,3})_0} &
V_{(\mathbf{1,1})_0} & V_{(\mathbf{3,2})_{-5/6}} & V_{(\mathbf{\bar 3,2})_{5/6}} \\ \hline
Z_n & 0 & 0 & 0 & y & -y \\
\hline
\end{array}
\end{equation}
Then the other representations in $E_7$ adjoint is decomposed in the following:
\begin{eqnarray}
(\mathbf{5,\bar 3})_2 &=& (H_C)_i\ [y-a_i]: (\mathbf{3,1})_{-1/3} + (H_u)_i\ [-a_i]:(\mathbf{1,2})_{1/2} \\
(\mathbf{\bar 5,3})_{-2} &=& (D^c)^i\ [-y+a_i]: (\mathbf{\bar 3,1})_{1/3}
+ L^i \ [a_i]: (\mathbf{1,2})_{-1/2} \\
(\mathbf{5,1})_{-3} &=& H_C \ [z+y]:(\mathbf{3,1})_{-1/3} + H_u \ [z]: (\mathbf{1,2})_{1/2} \\
(\mathbf{\bar 5,1})_3 &=& \bar H_C \ [-z-y]:(\mathbf{\bar 3,1})_{1/3}
+ H_d \ [-z]: (\mathbf{1,2})_{-1/2} \\
(\mathbf{10,\bar 3})_{-1} &=& Q_i \ [z+y-a_i]:(\mathbf{3,2})_{1/6}
+ U^c_i \ [z+2y-a_i]:(\mathbf{\bar 3,1})_{-2/3} \nonumber \\
&& + E^c_i \ [z-a_i]: (\mathbf{1,1})_1 \\
(\mathbf{\overline{10},3})_1 &=& (Q^c)^i \ [-z-y+a_i]:(\mathbf{\bar 3,2})_{-1/6}
+ U^i \ [-z-2y+a_i] :(\mathbf{3,1})_{2/3} \nonumber \\
&&+ E^i \ [-z+a_i]: (\mathbf{1,1})_{-1}  \\
(\mathbf{1,8})_0 + (\mathbf{1,1})_0 &=& S^i_j \ [a_i-a_j]: (\mathbf{1,1})_0
\end{eqnarray}
where the scripts $i,j$ denote $SU(3)$ indices and 
$Z_n$ charges for the corresponding vector superfields are given in the square brackets.
The automorphism condition is that $Q_i U^c_j (H_u)_k \epsilon^{ijk}$ is invariant:
$2z + 3y - a_1 - a_2 -a_3 \equiv 0$ (mod $n$).
Then choosing $l$ and $m$ appropriately,
the colored Higgs $H_C$ and $\bar H_C$ are all projected out from massless spectrum
and we can extract one massless family.
For example, we choose the parameters as
\begin{eqnarray}
&& z+y-a_1 \equiv 1 \equiv z-a_3, \\
&& z+2y -a_2 \equiv -\ell \equiv -z, \\
&& -a_3 \equiv -m \equiv -y+a_1 \equiv a_2, \\
&& z+y \neq \pm 1, \pm \ell, \pm m, \\
&& y-a_i \neq 1,-\ell,-m,
\end{eqnarray}
then
the bulk gauge interaction includes the Yukawa couplings for one family:
\begin{equation}
Q_1 U^c_2 (H_u)_3 + Q_1 (D^c)^1 H_d + E^c_2 L^2 H_d.
\end{equation}
We identify this one family as the third family,
then we find that all the three (conventional) gauge couplings of the Standard Model
and the third family  Yukawa couplings are unified.
%
This is an interesting example of the gauge--Yukawa unification where the
4D gauge group is the Standard Model with some additional 
$U(1)$ factors.

A similar model can be constructed from $E_8$ bulk symmetry.
In the $E_8$ model, Dirac neutrino Yukawa interaction can be also unified.

\subsection{Trinification : $SU(3)_c \times SU(3)_L \times SU(3)_R$}

The gauge--Higgs unification in $SU(3)^3$ gauge theory is briefly
mentioned in the Ref.\cite{Arkani-Hamed:2001tb}. The bulk
symmetry is $SU(9)$ and the $SU(9)$ symmetry is broken down to
$SU(3)^3 \times U(1)^2$. The $SU(9)$ adjoint is decomposed as
\begin{equation}
\mathbf{80} = \left(
\begin{array}{ccc}
 (\mathbf{8,1,1}) & (\mathbf{3,\bar 3,1}) & (\mathbf{3,1,\bar 3}) \\
 (\mathbf{\bar 3,3,1}) & (\mathbf{1,8,1}) & (\mathbf{1,3,\bar 3}) \\
 (\mathbf{\bar 3,1,3}) & (\mathbf{1,\bar 3,3}) & (\mathbf{1,1,8})
\end{array}
\right)
+ (\mathbf{1,1,1}) + (\mathbf{1,1,1}).
\end{equation}
We can extract the massless modes $(\mathbf{3,\bar 3,1})$,
$(\mathbf{\bar 3,1,3})$, $(\mathbf{1,3,\bar 3})$
from three chiral multiplets,
and those can be identified as left-handed quark $Q_L$, right-handed quark
$Q_R^c$ and Higgs field $H$ in the trinification model, respectively.
The trilinear bulk gauge interaction includes the Yukawa interaction
$Q_L Q_R^c H$.
In this assignment, there is no lepton multiplet in the gauge multiplet.
Of course, we can assign that $(\mathbf{1,\bar 3,3})$
is lepton representation, but lepton Yukawa coupling is not included
in the bulk interactions.

We can also construct a trinification model from $E_7$ bulk symmetry:
$E_7$ has a breaking branch to trinification as
\begin{equation}
E_7 \rightarrow SU(6) \times SU(3) \rightarrow SU(3)^3 \times U(1).
\end{equation}
In this case, lepton Yukawa couplings are not included in the bulk gauge interactions, either.
In the $E_8$ bulk symmetry,
we can make a trinification with lepton Yukawa couplings
arising from bulk gauge couplings:
$E_8$ has a breaking branch to trinification as
\begin{equation}
E_8 \rightarrow SU(9) \rightarrow SU(3)^3 \times U(1)^2.
\end{equation}
The adjoint of $E_8$ is decomposed as
\begin{equation}
\mathbf{248} = \mathbf{80}+ \mathbf{84} + \overline{\mathbf{84}},
\end{equation}
and $\mathbf{80}$ includes 
$(\mathbf{3,\bar 3,1})$, $(\mathbf{\bar 3, 1,3})$, $(\mathbf{1,3,\bar 3})_H$
for both left- and right-handed quarks and also Higgs
as we have seen above,
and $\mathbf{84}$ and $\overline{\mathbf{84}}$ include lepton multiplets
$(\mathbf{1,3,\bar 3})_{\mathbf{84}}$ and $(\mathbf{1,3,\bar 3})_{\overline{\mathbf{84}}}$.
The Yukawa couplings for quarks are included in the bulk interaction
$\mathbf{80}\cdot\mathbf{80}\cdot\mathbf{80}$,
and the lepton Yukawa couplings are in the 
\begin{equation}
\mathbf{80}\cdot\mathbf{84}\cdot\overline{\mathbf{84}}\supset
(\mathbf{1,3,\bar 3})_{\mathbf{84}} (\mathbf{1,3,\bar 3})_{\overline{\mathbf{84}}}
(\mathbf{1,3,\bar 3})_H.
\label{lepton-yukawa}
\end{equation}
The left- and right-chiral quarks are separated in the different multiplets,
so that the gauge and Yukawa couplings can be unified.
For the lepton, however, since both chirality of leptons 
are unified in one multiplet $(\mathbf{1,3,\bar 3})$,
the Yukawa coupling Eq.(\ref{lepton-yukawa}) includes Yukawa couplings for two families with
degenerate couplings.
So, as we have seen previously in the case $SO(10)$, 
the Yukawa coupling constant for the lepton is not unified completely.
If we break $SU(3)_R$ symmetry by orbifold,
we can make the situation
where $(\mathbf{1,3,\bar 3})_{\mathbf{84}}$ includes only left-handed lepton
and $(\mathbf{1,3,\bar 3})_{\overline{\mathbf{84}}}$ includes only right-handed lepton.
In that case, lepton Yukawa couplings can be also unified to the  gauge and quark
Yukawa couplings.

We comment that the 4D left--right symmetric model
whose gauge group is $SU(3)_c \times SU(2)_L \times SU(2)_R \times U(1)_{B-L}$
can be also constructed by using this trinification branch of $E_8$ or
Pati--Salam branch of $SO(16)$ bulk symmetry.

\subsection{Classification of the Models by Bulk Gauge Symmetry}

We have classified gauge--Higgs (and matter) 
unified models by 4D symmetry after orbifold breaking
in the previous subsections.
In this subsection, we classify the unification models by bulk gauge symmetry.
We only consider the case where the bulk gauge symmetry is a simple group.
We have basically dealt with the possibility that
three Yukawa (top--bottom--tau) couplings are unified to the gauge coupling at
GUT scale, whereas
we can also obtain the situation where only top-Yukawa coupling is unified to 
gauge coupling if we choose smaller rank of bulk gauge symmetry.
We call the situation ``partial gauge--Yukawa unification".

In the $SU$ series, minimal choice to obtain gauge--Higgs unification is $SU(6)$.
We have already seen the gauge--Higgs unification in
the 5D $SU(6)$ model
in subsection 3.3.
We can construct the 6D $SU(6)$ model.
The $SU(6)$ is broken down to $SU(3)_c \times SU(2)_L \times U(1)^2$
by orbifold condition,
and the adjoint representation is decomposed as
\begin{equation}
\mathbf{35}= \left( \begin{array}{ccc}
                     (\mathbf{8,1})_{0,0} & (\mathbf{3,2})_{1,0} & (\mathbf{3,1})_{1,3} \\
                     (\mathbf{\bar 3,2})_{-1,0} & (\mathbf{1,3})_{0,0} & (\mathbf{1,2})_{0,3} \\
                     (\mathbf{\bar 3,1})_{-1,-3} & (\mathbf{1,2})_{0,-3} & (\mathbf{1,1})_{0,0}
                    \end{array}
             \right) + (\mathbf{1,1})_{0,0} \ .
\end{equation}
The $U(1)^2$ charges $(X_1,X_2)$ are given in the subscripts.
The hypercharge is given as $Y= (X_1 + X_2)/6$.
The quark doublet and right-handed up-type quark can be included in the gauge multiplet,
and the trilinear bulk gauge interaction includes up-type Yukawa coupling
\begin{equation}
(\mathbf{3,2})_{1,0} \ (\mathbf{\bar 3,1})_{-1,-3} \ (\mathbf{1,2})_{0,3} .
\end{equation}
So the gauge couplings and top-Yukawa coupling can be unified at compactification scale.
However, the weak mixing angle prediction at unified scale is not usual $3/8$ but $3/4$,
and thus the gauge coupling unification is non-trivial in the 6D $SU(6)$ model.
We can also construct 6D $SU(7)$ model.
The orbifold condition breaks $SU(7)$ down to $SU(3)_c \times SU(2)_L \times SU(2)_R \times U(1)^2$.
So we can make left--right symmetric model,
and the both left- and right-handed quarks can be included in the gauge multiplet.
We can see that the trilinear bulk gauge coupling contains quark Yukawa couplings,
but again the gauge coupling unification is non-trivial.
Going up to $SU(8)$, we can construct Pati--Salam model
with both quark and lepton  included in the gauge multiplet
as we have seen in subsection 3.2.
All the gauge and Yukawa couplings can be unified in the bulk $SU(8)$ model.

In the $SO$ series, minimal choice to obtain gauge--Higgs unification is
$SO(11)$ \cite{Cosme:2003cq}.
We have seen the example as 5D $SO(12)$ model in subsection 3.1.
To obtain gauge--Higgs and also matter unification,
we can construct 6D bulk $SO(12)$ model.
The $SO(12)$ is broken down as
\begin{equation}
SO(12) \rightarrow \left\{
                    \begin{array}{c}
                     SU(6) \times U(1) \\
                     SO(10) \times U(1) 
                    \end{array}
                   \right\}
       \rightarrow SU(5)\times U(1)^2.
\end{equation}
The adjoint representation is decomposed as
\begin{eqnarray}
\mathbf{66} &=& \mathbf{35}_0 + \mathbf{1}_0 + \mathbf{15}_2 + \mathbf{\overline{15}}_{-2}  \qquad
({\rm under} \ SU(6) \times U(1)) \nonumber \\
&=&   \mathbf{24}_{0,0} + \mathbf{5}_{0,3} + \mathbf{\bar 5}_{0,-3} + \mathbf{1}_{0,0} + \mathbf{1}_{0,0}
 +  \mathbf{10}_{2,1} + \mathbf{5}_{2,-2} + \mathbf{\overline{10}}_{-2,-1} +\mathbf{\bar 5}_{-2,2}.
\end{eqnarray}
The trilinear bulk interaction includes
\begin{equation}
\mathbf{10}_{2,1} \ \mathbf{\bar 5}_{-2,2}\ \mathbf{\bar 5}_{0,-3}\ ,
\end{equation}
and this can be down-type Yukawa coupling.
If we consider flipped-type $SU(5)$, this can be up-type Yukawa coupling.
Thus, in the $SO(12)$ model, matter can be partially unified in the gauge multiplets,
and also Yukawa couplings are partially unified to the gauge couplings.
The $SO(16)$ bulk model is investigated in Ref.\cite{Gogoladze:2003yw}.
In the $SO(16)$ model, two families of matters can be unified in the gauge multiplet.

For the $E$ series, 
the gauge--Higgs unification in 5D model is
considered in the Ref.\cite{Haba:2002vc}.
We have already seen gauge--Higgs--matter unification 
in the 6D $E_7$ model
in the subsection 3.3 and 3.4
and the 6D 
$E_8$ model can be also constructed (for example, trinification model in subsection 3.5).
The 6D $E_7$ and $E_8$ model, both quark and lepton can be unified into
the gauge multiplet.
We can construct partially matter unification in $E_6$ model.
The bulk $E_6$ symmetry can be broken by orbifold condition
and one of the branch is $SU(6) \times SU(2)$.
The adjoint representation is decomposed as
\begin{equation}
\mathbf{78} = (\mathbf{35,1}) + (\mathbf{1,8}) + (\mathbf{20,2}).
\end{equation}
The trilinear bulk gauge interaction contains $\mathbf{(20,2)\ (20,2)\ (35,1)}$
which includes up-type Yukawa coupling $\mathbf{10}_1 \mathbf{10}_2 \mathbf{5}$ 
in $SU(5)$ decomposition, where the subscripts 1,2 are $SU(2)$ indices. 
Since two eigenmasses are degenerate in the bulk Yukawa coupling in this case,
we have to deal with the degeneracy as we have seen previously.

\section{Remarks on the Conditions of Gauge--Yukawa Unification}

The unification of gauge and Higgs in higher dimensions
leads us naturally to the possibility that the 
Yukawa interactions originate from gauge interactions.
Here, we make some remarks on the unification of
gauge and Yukawa couplings.

Precisely speaking,
the unification of gauge and Yukawa couplings  can be realized
in the higher dimensional Lagrangian
and the unifying constants are higher dimensional coupling constants.
So, it does not mean directly the unification of gauge and Yukawa
couplings in 4D.
We need to have some conditions to realize the unification,
and at most case, the coupling unification will be zero-th order relations.
Nonetheless, this possibility of gauge--Yukawa unification seems
quite important since origin of Yukawa interaction
is one of the mystery in the Standard Model.

The conditions to realize the unification of gauge--Yukawa constants
are following.
1) The brane-localized gauge and Yukawa interactions
and their threshold corrections can be negligible.
2) The zero modes of fermions are not localized
at different points on the orbifold space.
3) The 4D fields are not largely mixed with other brane-localized fields.

The condition 1) is the same condition to realize
gauge coupling unification in higher dimensional GUT.
Assuming the large volume suppression
of extra dimensions,
the brane-localized interaction can be negligible.
The threshold effects are model-dependent basically,
and those corrections can be negligible except for the strong coupling constants.

Since the unification of gauge and Yukawa couplings is realized
in higher dimensional Lagrangian,
we have to integrate the extra dimensions to obtain the relation of the 4D couplings.
The unification relations of the 4D couplings hold
if we assume that
the 4D couplings are just higher dimensional coupling multiplying
volume factor of extra dimensions.
If zero modes are localized in extra dimensions, this situation is not necessary
satisfied.
So we assume that third generation is not localized in extra dimensions.

In the $SO(10)$ and $SU(5)$ case in subsection 3.1 and 3.3, the condition 3) is not satisfied
and the gauge and Yukawa couplings  are not unified in general.

Other correction coming from K\"ahler potential can be considered.
We expect that such corrections at the
unification scale are less than 5\% level which is $O(M_{\rm GUT}/M_{\rm Planck})$.
%
%
The detailed numerical study of gauge--Yukawa unification has been
performed and we have some predictions about $\tan \beta$
and the threshold corrections for supersymmetry breaking \cite{Gogoladze:2003pp}.

In this paper, we have concentrated on the third family unification.
In principle, other families can be unified in our picture.
For example,
the second family is also unified in 6D model with $SO(16)$
bulk symmetry \cite{Gogoladze:2003yw}.
Other families may be brane-localized fields.
In fact, in our construction,
there are gauge anomalies for extra $U(1)$ symmetries
if we only have zero modes of bulk families
and we need brane families to cancel the anomalies.
Then the Yukawa coupling for other families are naturally
small because of the volume suppression of extra dimensions.
We can also use Froggatt--Nielsen like mechanism \cite{Froggatt:1978nt} to
make hierarchical structure of quark and lepton mass matrices for the first and second families
using the extra $U(1)$ factors.

\section{Conclusion}

The origin of families is one of the mystery in particle physics.
The quark and lepton masses and their mixing angles are just parameters
in the Standard Model.
The Higgs scalar fields are also one of the mysteries in the Standard Model.
The idea of gauge--Yukawa unification in higher dimensions
may give us an interesting approach
to consider such mysteries in the unified picture.

We need to have some assumption
to  realize the unification of the gauge and Yukawa couplings
as we mentioned in section 4.
In the case where the 4D symmetry
is $SO(10)$ or $SU(5)$,
the gauge and Yukawa couplings  are not unified in general.
It is interesting that those couplings  can be unified
in the case where the 4D symmetry is
Pati--Salam or Standard Model.
Because of the bulk symmetry, the gauge couplings
can be unified at grand unified scale as well as
Yukawa couplings for third family.
The numerical predictions are in good agreement with
experimental data \cite{Gogoladze:2003ci}
and we have predictions for $\tan \beta$ and relations between 
threshold corrections for the supersymmetry breaking \cite{Gogoladze:2003pp}.
At the same time, the colored Higgs fields can be projected out
by orbifold conditions.
So, the scenario of gauge--Yukawa unification is
compatible with standard unified pictures.
In fact, in order to make gauge--Higgs unification
we need to extend gauge group in the higher dimensions,
and this can be the motivation to consider the grand unified theories
in higher dimensions.
We emphasize that
the unification of gauge and Yukawa interaction
is not a special situation in the case where
gauge and Higgs fields are unified in higher dimensional models
and quarks and leptons are zero modes of bulk fermions.
The gauge covariant forms of Kaluza--Klein mass terms
naturally include the Yukawa interactions.

\section*{Acknowledgments}

We thank K.S. Babu, W.A. Bardeen, B. Dutta, C. Kolda, T. Li, J.D. Lykken,  R.N. Mohapatra,
H.P. Nilles, Y. Nomura, J.C. Pati, S. Raby, and E. Witten for useful
discussions and comments.
The work of I.G. was supported in part by 
the National Science Foundation under grant PHY00-98791 and
by the U.S.~Department of Energy under grant Numbers DE-FG03-98ER-41076
and DE-FG-02-01ER-45684. The work of Y.M. was supported  by the Natural Sciences and
Engineering Research Council of Canada,
and the work of S.N. was supported in part by US Department of Energy Grant Numbers DE-FG03-98ER-41076
and DE-FG-02-01ER-45684.

\end{document}